\documentclass[]{spie}  
\usepackage[]{graphicx}
\usepackage{subfig,booktabs,ctable}
\usepackage{epstopdf}
\usepackage{multirow}

\title{PCA Method for Automated Detection of Mispronounced Words}

\author{Zhenhao Ge, Sudhendu R. Sharma, Mark J.T. Smith
\skiplinehalf
School of Electrical and Computer Engineering \\
Purdue University, West Lafayette, IN, USA, 47907
}

\authorinfo{Further author information: (Send correspondence to Zhenhao Ge)\\
Zhenhao Ge: E-mail: zge@purdue.edu, Telephone: 1 317 457 9348\\
Sudhendu R. Sharma: E-mail: sharmasr@purdue.edu, Telephone: 1 405 334 9992}

\begin{document}
\maketitle 

\begin{abstract}
This paper presents a method for detecting mispronunciations with the aim of improving Computer Assisted Language Learning (CALL) tools used by foreign language learners. The algorithm is based on Principle Component Analysis (PCA). It is hierarchical with each successive step refining the estimate to classify the test word as being either mispronounced or correct. Preprocessing before detection, like normalization and time-scale modification, is implemented to guarantee uniformity of the feature vectors input to the detection system. The performance using various features including spectrograms and Mel-Frequency Cepstral Coefficients (MFCCs) are compared and evaluated. Best results were obtained using MFCCs, achieving up to 99\% accuracy in word verification and 93\% in native/non-native classification. Compared with Hidden Markov Models (HMMs) which are used pervasively in recognition application, this particular approach is computational efficient and effective when training data is limited. 
\end{abstract}


\section{INTRODUCTION}
\label{sec:intro}  

With the advent of technology, many Computer Assisted Language Learning (CALL) and Computer Assisted Pronunciation Training (CAPT) tools are available in the market to help people learn a new language. Learning a new language can be a very difficult. Without proper feedback, it can be very frustrating. Most CALL tools focus on teaching new words or sentences using a listen-and-repeat procedure. These tools use modern automatic speech recognition (ASR) algorithms and provide visual aids such as spectrograms and waveforms as feedback. However, these tools are not geared towards identifying specific mispronunciations and consequently are a poor substitute for a human instructor. The work discussed in this paper focuses on mispronunciation detection.

In the last two decades, a significant amount of research has been carried out in the field of mispronunciation detection.  The majority of the mispronunciation detection studies \cite{Franco,Ronen,Wang,Harrison,Harrison1,Wei,Qian} have extensively used ASRs based on statistical models, such as Hidden Markov Models (HMM). However, as Garc\'{\i}a-Moral et al. \cite{Garcia-Moral} point out, for HMM based systems ``large databases are required to warrant relevant and statistically reliable results". For mispronunciation detection, these systems have to train different models for correct pronunciations and mispronunciations. Recent works by Wang et al. \cite{Wang}, and Harrison et al. \cite{Harrison1} have shown improvement in detection accuracy, but these systems had to use extended recognition network by making use of cross-language phonological rules, i.e. rules that dictate how a learner's first language affects his/her pronunciation of a second language. The extended recognition networks require additional models to be trained in order to account for the cross-language rules. The use of large database and a large number of models makes HMM-based systems computationally complex and costly. Furthermore, performance often suffers when the training set is limited.

In this paper, we present a mispronunciation detection system based on Principal Component Analysis (PCA) and show it is computationally efficient compared to HMM's and performs well with limited training data.
The approach used in this paper for detecting mispronunciation is to classify pronunciations, at both the word-level and syllable-level, as either acceptable or unacceptable.

\section{PCA METHOD FOR PATTEN RECOGNITION}

PCA is a well-known method that has been successfully employed in high-dimensional recognition problems.  One of the popular applications of PCA is for face recognition which has been studied extensively over the past 20 years \cite{Sirovich,Kirby,Belhumeur,Mandal,Turk&Pentland}. The fundamental idea behind PCA is encoding the most relevant information that distinguishes one pattern from another and eliminating dimensions with less information to reduce the computational cost \cite{Shlens}.

The mispronunciation detection work in this paper shares much in common with face recognition. Mispronunciation detection can be thought of as a pattern recognition problem, just like face recognition. Furthermore, Mel-Frequency Cepstral Coefficients (MFCCs) and spectrograms, which are the features that we are investigating, are also of high dimension. The similarities between mispronunciation detection and face recognition motivated our incorporating PCA-based face recognition techniques into our work. In the next section, we first review the application of PCA in face recognition, and then discuss the procedures of the PCA-based mispronunciation detection.

\subsection{Review of Face Recognition using PCA  }

The PCA approach for face recognition requires all data (face images), training data $\mathcal{D}_{\rm{train}}$, and testing data $\mathcal{D}_{\rm{test}}$ from multiple classes to have the same size (say $N_1 \times N_2$). The method tends to work best when the faces in all the images are located at the same position within the image. These data matrices are then vectorized to $N_1N_2$-dimensional column vectors. When implementing a PCA-based detection/classification system, like face recognition, given $M$  vectorized training faces $\Gamma_1,\Gamma_2,\dots,\Gamma_M (M \ll N_1N_2)$ as the whole set of $\mathcal{D}_\mathrm{train}$, let $\Psi = \frac{1}{M}\sum_{i=1}^M\Gamma_i $ be the mean face and let $\Phi_i = \Gamma_i - \Psi$ be the mean-shifted faces which form $A=[\Phi_1,\Phi_2,...,\Phi_M]$. $C=AA^T$ is the covariance matrix of $A$, and can supply up to $M$ eigenvectors to represent the eigenspace of $A$. Each face in $A$ is a linear combination of these eigenvectors. If selecting the most significant  $M^\prime$ eigenvectors to form a sub-eigenspace of training data $U=[u_1,u_2,\dots,u_{M^\prime}]$ ($N_1N_2 \times M^\prime$), the representation of $\Phi_i$ in the $M^\prime$-dimensional eigenspace is $\Omega_i=U^T \Phi_i$ and the projection of $\Phi_i$ in the eigenspace is $\hat{\Phi}_i=U\Omega_i$. The dimension of data is reduced from $N_1N_2$ to $M^\prime$ while preserving the most relevant information to distinguish faces.

\begin{figure}
 \centering
   \includegraphics[scale=0.7]{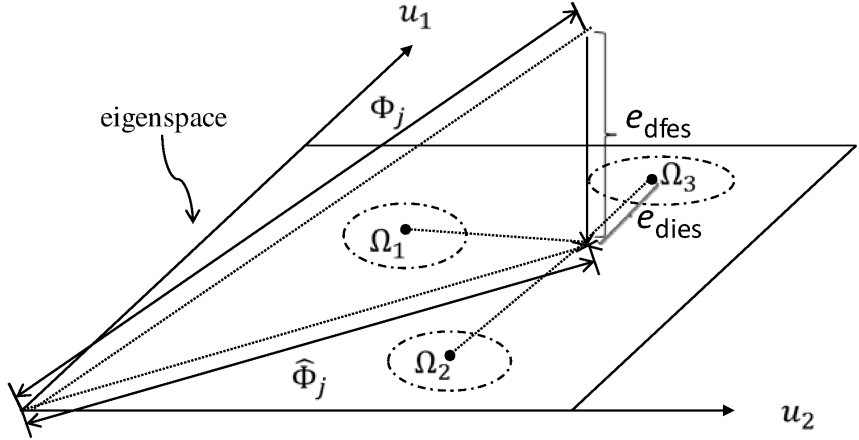}
   \caption{A simplified version of eigenspace to illustrate $e_\mathrm{dfes}$ and $e_\mathrm{dies}$  \label{eigenspace}}		
\end{figure}

After the eigenspace $U$ for the training data $A$ is computed, the following two steps are used for face recognition\cite{Turk&Pentland}: 

\begin{enumerate}
\item Face detection: compute the Euclidean distance from test image $\Phi_j$ to its projection $\hat{\Phi}_j$ onto the eigenspace $U$,
\begin{equation} \label{dfes}
	e_\mathrm{dfes} = \Vert\Phi_j-\hat{\Phi}_j\Vert ~ \mbox{,} 
\end{equation}
where $e_\mathrm{dfes}$ is called ``distance from eigenspace". If $e_\mathrm{dfes} < T_d$, where $T_d$ is a detection threshold, the test image is verified to be a face and step 2 below is used for face classification. 

\item Face classification: compute the minimum Euclidean distance between $\Omega_j$ and $\Omega_k$ in Equation (\ref{dies}), where $\Omega_j$ is the eigenspace representation of the test face $\Phi_j$ and $\Omega_k$ is the averaged eigenspace representation of faces in class $k$, $k=1,2,\dots,K$, where $K$ is the number of classes. 
\begin{equation} \label{dies}
	e_\mathrm{dies} = \min{\Vert\Omega_j-\Omega_k\Vert}\qquad k=1,2,\dots,K 
\end{equation}  
where $e_\mathrm{dies}$ is called ``distance within eigenspace". If $e_\mathrm{dies}<T_c$, where $T_c$ is the classification threshold, then the test image is classified as a face belonging to that class; otherwise, a new class of faces is claimed. Figure \ref{eigenspace} illustrates the difference between $e_\mathrm{dfes}$ and $e_\mathrm{dies}$.
\end{enumerate}

\subsection{Procedure for Mispronunciation Detection Using PCA}

In this work, the training and testing data consist of two disjointed classes of speech samples. The first class consists of samples from native speakers, all of which have correct pronunciations. The second class contains samples from non-native speakers with possible mispronunciations. Given an input test sample, the procedure used for mispronunciation detection can be divided into 3 steps:

\begin{enumerate}
\item Word verification: compute $e_\mathrm{dfes}$ from the test sample $\Phi_j$ to the eigenspace $U_\mathrm{All}$ constructed from all samples in  $\mathcal{D}_{\mathrm{All.train}}$, where $\mathcal{D}_{\mathrm{All.train}}$ includes all native $\mathcal{D}_{\mathrm{N.train}}$ and non-native $\mathcal{D}_{\mathrm{NN.train}}$ samples. If $e_\mathrm{dfes}<T_d$, where $T_d$ is the threshold for word verification, then the test sample is considered ``verified" and we proceed to step 2. Otherwise the test sample is ``rejected" and the detection process stops. 

\item Native/non-native (N/NN) classification: compute $e_\mathrm{dfes}$ from the test sample $\Phi_j$ to the eigenspace $U_\mathrm{N}$ constructed only from  $\mathcal{D}_{\mathrm{N.train}}$. If $e_\mathrm{dfes}<T_c$, then classify the test sample as native and stop. Otherwise, classify it as non-native and proceed to step 3.

\item Syllable-level mispronunciation detection: Divide the non-native test samples into syllables $\Phi_{jk},k=1,2,\dots,K$, where $K$ is the number of syllables for that test sample. Similarly divide native samples $\mathcal{D}_{\mathrm{N.train}}$ into syllables and call the $k^{\mathrm{th}}$ syllable ${D}_{\mathrm{N}k.\mathrm{train}}$. For each test syllable $\Phi_{jk}$, compute its $e_\mathrm{dfes}$ to the eigenspace $U_{\mathrm{N}k}$  constructed from $k^{\mathrm{th}}$ syllable of native samples $\mathcal{D}_{\mathrm{N}k.\mathrm{train}}$. If $e_\mathrm{dfes} > T_k$, then classify the test syllable as mispronounced, otherwise classify it as correctly pronounced.  

\end{enumerate}

\begin{figure}
  \centering
  \subfloat[case 1]{
  \label{case1}\includegraphics[width=0.45\textwidth]{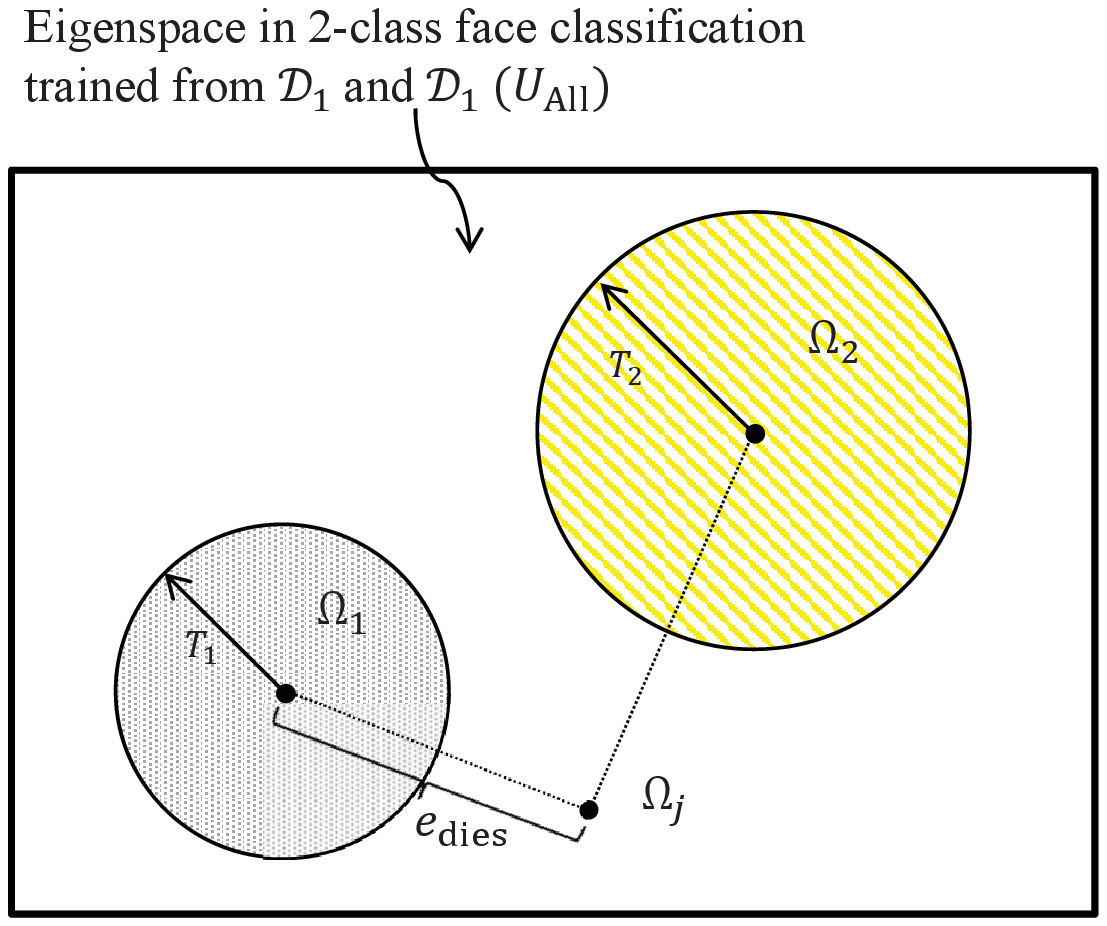}}                
  \subfloat[case 2]{
  \label{case2}\includegraphics[width=0.45\textwidth]{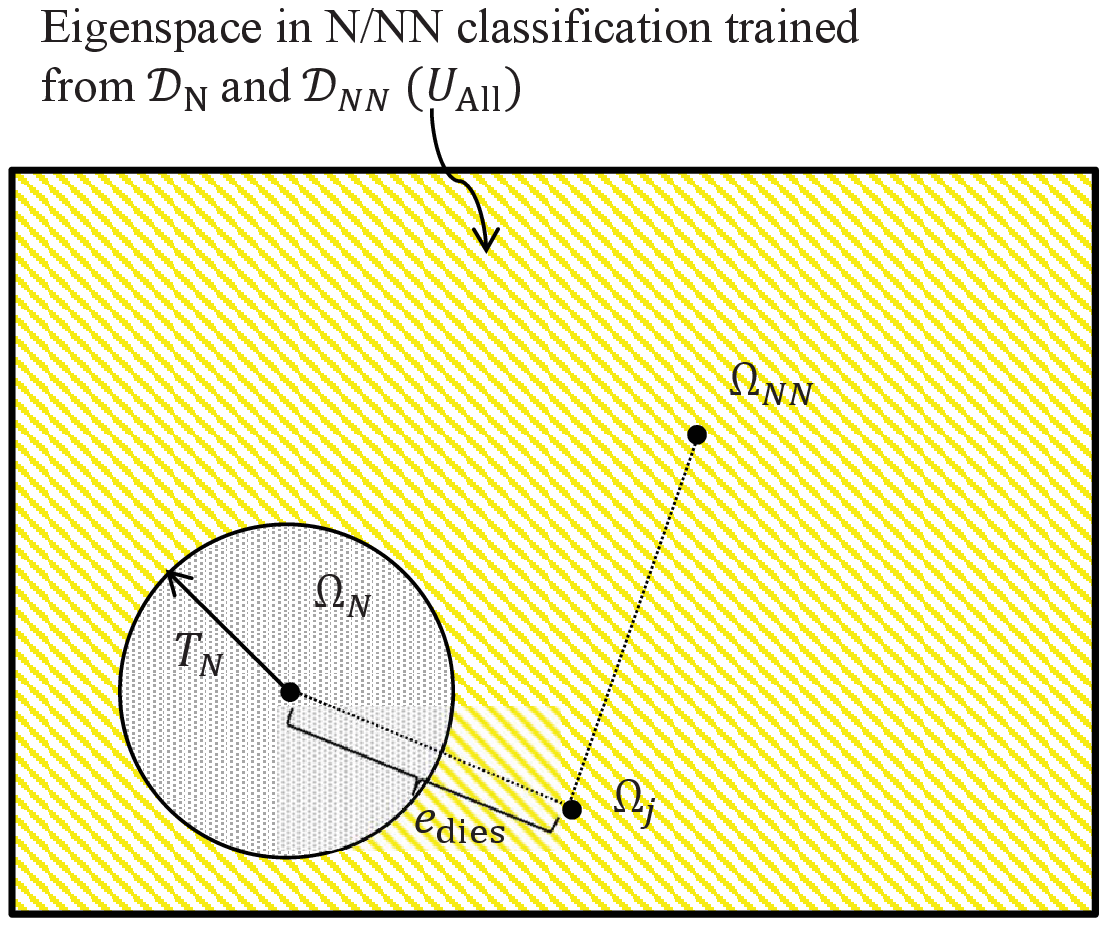}}
  \caption{Possible distributions of native and non-native samples in the eigenspace $U_\mathrm{All}$}
  \label{cases}
\end{figure}

Word verification is very similar to face detection, since both cases use the distance metric $e_\mathrm{dfes}$ of the eigenspace $U_\mathrm{All}$ which is constructed from all classes of samples. However, N/NN classification is different from face classification. The main difference is that in face classification $e_\mathrm{dies}$ is used as the metric to determine the class of the ``test" face, where as in mispronunciation detection we use $e_\mathrm{dfes}$ as the metric to do the N/NN classification. In face classfication, the various classes of faces are well defined. If $e_\mathrm{dies}$ of a ``test face" lies within the predefined threshold of some class $k$, then it is classified as face $k$. Otherwise, it is claimed to be a ``new face" that does not match any of the faces. The ``new face" scenario for face classification is shown in Figure \ref{case1}, where $\Omega_j$ is the ``test" face. However, in N/NN classification, there are only two classes and any region outside the native class belongs to the non-native class. So the misclassification shown in Figure \ref{case2} will occur if $e_\mathrm{dies}$ is used to determine the class of $\Omega_j$. In Figure \ref{case2}, $\Omega_j$ is in the non-native class but it is closer to native class. Thus, instead of using distance $e_\mathrm{dies}$ within eigenspace $U_\mathrm{All}$, N/NN classification uses $e_\mathrm{dfes}$ to measure the distance from the eigenspace $U_\mathrm{N}$, which is a sub-eigenspace of $U_\mathrm{All}$.

The syllable-level mispronunciation detection is an extension of the N/NN classification applied to syllables. Each ``test syllable" is treated like a ``test word" in step 2, and mispronunciation can be detected in the syllables using the same N/NN classification approach.   

\section{SYSTEM DESIGN AND IMPLEMENTATION}

Having reviewed the  application of PCA in face recognition and how it can be applied to mispronunciation detection, the following section will discuss the implementation of PCA-based mispronunciation detection in detail, including database construction, data pre-processing, feature selection, eigenspace training, and detection threshold optimization. 

\subsection{Database Construction}

Although the algorithm is not language specific, Spanish was chosen for this work. The database used in this work is relatively small and contains only 10 Spanish words listed in Table \ref{wordlist}. These words were selected by language experts to cover a variety of common mispronunciations exhibited by American speakers learning Spanish. 
There were 13 male speakers, 7 of them native speakers and 6 non-native. Each speaker repeated each of the 10 words 5 times. 

Because of the limited size of the database, the Leave-One-Out method was used for training and testing and is further discussed in Section 3.4.

\begin{table}[h]
\centering
\caption{List of Spanish words and syllables}
\label{wordlist}
\begin{tabular}{@{} cll|cll @{}} \toprule%
\multicolumn{1}{>{\bfseries}c}{No.} & 
\multicolumn{1}{>{\bfseries}l}{Words} & 
\multicolumn{1}{>{\bfseries}l}{Syllables} & 
\multicolumn{1}{>{\bfseries}c}{No.} & 
\multicolumn{1}{>{\bfseries}l}{Words} & 
\multicolumn{1}{>{\bfseries}l}{Syllables}\\\midrule 
1 & {\tt jamaica} & /ja/, /mai/, /ca/ & 6 & {\tt tordurados} & /tor/, /tu/, /ra/, /do/, /s/ \\
2 & {\tt tres} & /tr/, /e/, /s/ & 7 & {\tt accidente} & /ac/, /ci/, /den/, /te/ \\
3 & {\tt gemelas} & /ge/, /me/, /la/, /s/ & 8 & {\tt construccion} & /cons/, /truc/, /cion/ \\
4 & {\tt hierro} & /hie/, /rro/ & 9 & {\tt puertorriquena} & /puer/, /tor/, /ri/, /que/, /na/ \\
5 & {\tt pala} & /pa/, /la/ & 10 & {\tt aire} & /ai/, /re/ \\\bottomrule
\vspace{-0.2 in}
\end{tabular}
\end{table} 

\subsection{Data Pre-processing and Feature Extraction}
    
The PCA method requires centralized and uniform-size input features for training and testing. For centralization, periods of silence before and after the actual speech segment were removed using voiced/unvoiced detection. To make all samples uniform in size, the samples were time-scale modified to match the average duration of the training data. Furthermore, to improve the detection performance, the background noise was suppressed and the amplitude of each sample was normalized to unity.

Spectrograms and MFCCs were chosen as input features to the detection system. These features were computed for frames with window size 25ms (using a Hamming window) and 15 ms overlap between frames. The spectrogram feature space is 50-dimensional with each dimension spanning 320 Hz to 16 KHz. The MFCC feature space was 13-dimensional representing the first 13 Mel-scale cepstral coefficients.

\subsection{Eigenspace Training and Detection Threshold Optimization}
\label{system training}

As discussed in Section 2.2, these three steps are: (a) word verification; (b) N/NN classification; and (c) syllable-level mispronunciation detection. The detection system runs on a word-by-word basis. For each word, $W_{i^*}$, $i^*=1,2,...,10$, each step (a), (b), and (c) undergoes the following 3 phases of training.
 
\begin{itemize}
\item Phase 1: Train to determine eigenspace $U$.
\item Phase 2: Compute two sets of distances $e_\mathrm{dfes}^1$ and $e_\mathrm{dfes}^2$, where $e_\mathrm{dfes}^1$ corresponds to the distances from ``class 1" samples to the eigenspace $U$ and $e_\mathrm{dfes}^2$ corresponds to the distance from ``class 2" samples to $U$.
\item Phase 3: Find an optimal detection threshold $T$ that separates these two sets of distances.
\end{itemize}

\begin{figure}[h]
\begin{minipage}[t]{0.49\linewidth} 
\centering
\captionsetup{width=0.9\textwidth} 
\includegraphics[height=6.5cm]{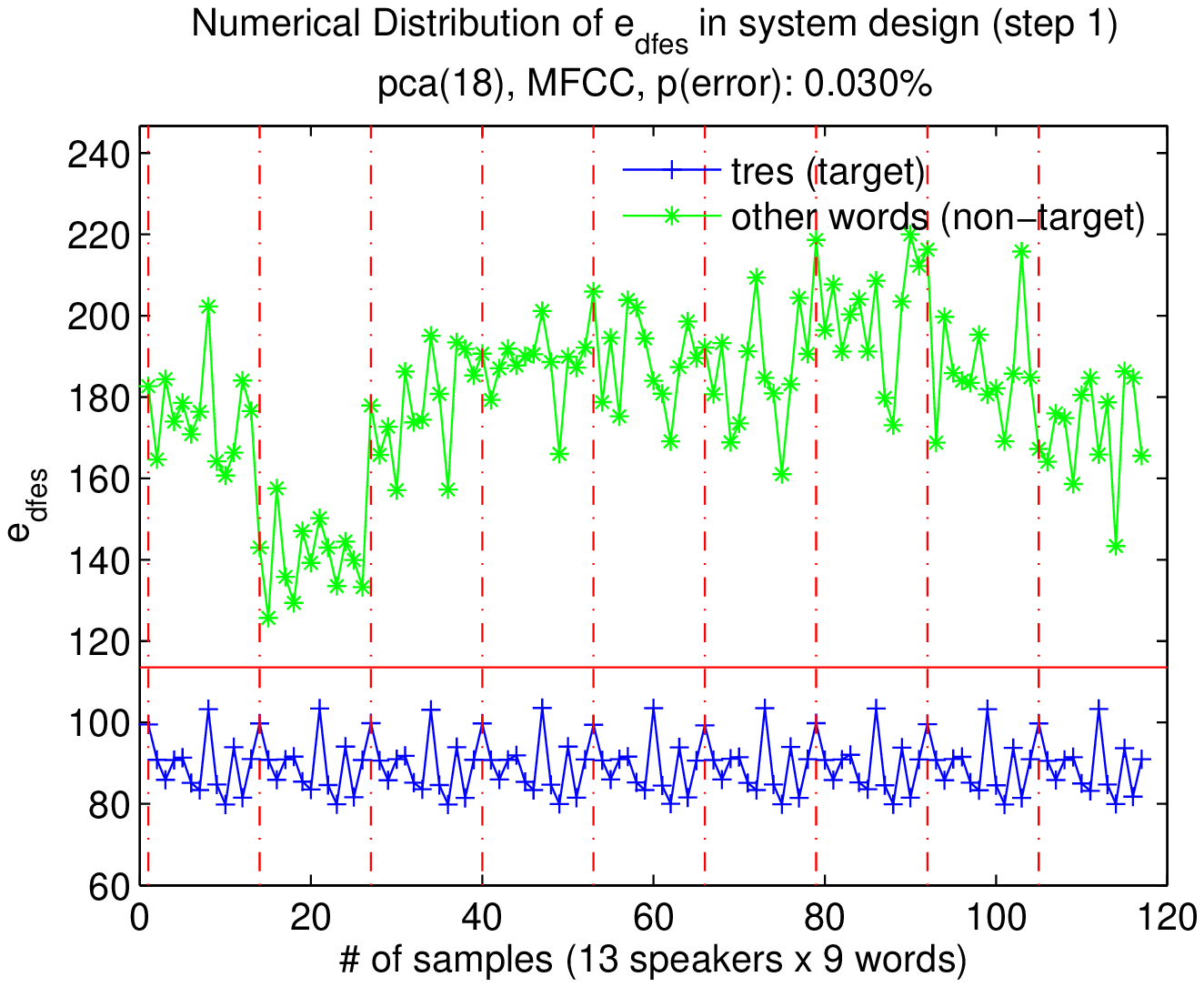}
\caption[width=0.25\textwidth]{Numerical distribution of class 1 and class 2 data in word verification (target word: {\tt tres}, feature: MFCCs )}  
\label{wdpcamfcc_numerical}
\end{minipage}
\hfill
\begin{minipage}[t]{0.49\linewidth}
\centering
\captionsetup{width=0.9\textwidth} 
\includegraphics[height=6.5cm]{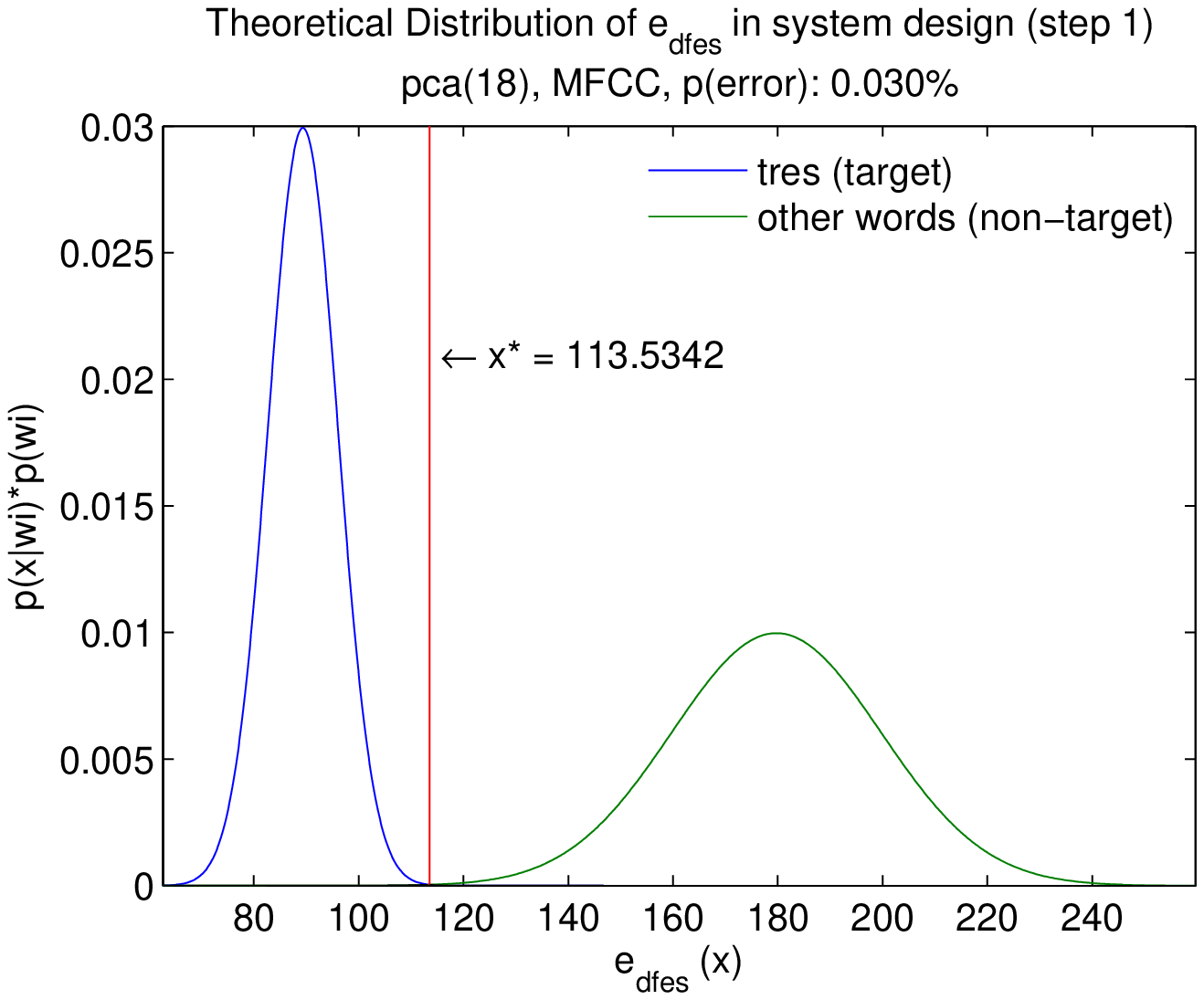}
\caption{Theoretical distribution of class 1 and class 2 data in word verification (target word: {\tt tres}, feature: MFCCs )}
\label{wdpcamfcc_theoretical}
\end{minipage}
\end{figure}

Even though the 3 training phases are the same for the 3 steps (a), (b), and (c), each step has a different trained eigenspace, uses different class 1 and class 2 data, and generates different thresholds. Table \ref{comparison} summarizes the difference in the 3 training phases of the 3 steps. In Table \ref{comparison}, the data used in training to obtain the eigenspace (2$^{\mathrm{nd}}$ row) and to compute $e_\mathrm{dfes}^1$ (3$^{\mathrm{rd}}$ row) are slightly different. Even though they share the same notation in the table, they represent two disjoint subsets within the class 1 data. Furthermore, different eigenspaces are trained using the Leave-One-Out approach. In this approach, samples corresponding to one speaker are left out. The samples from remaining speakers are used to train the eigenspace, and the samples from the left out speaker are used to find $e_\mathrm{dfes}^1$. This is repeated for all speakers belonging to class 1. As mentioned, this approach leads to several trained eigenspaces. The $e_\mathrm{dies}^2$ distances are then computed for class 2 data to those different trained eigenspaces and averaged for each speaker in class 2. The distributions formed by $e_\mathrm{dfes}^1$ and $e_\mathrm{dfes}^2$, an example shown in Figure \ref{wdpcamfcc_numerical}, are then used to find the optimal threshold. By convention \cite{PVCA@website}, the dimension of the eigenspace in each step is determined by selecting eigenvectors that represent 80\% variance of the total principle components. For example, the dimension of the eigenspace used to determine $e_\mathrm{dfes}$ in Figure \ref{wdpcamfcc_numerical} is 18. 

\begin{table}
\begin{center}
\caption{Data and eigenspace comparison of 3 mispronunciation detection steps}
\label{comparison}
\begin{tabularx}{0.8\linewidth}%
{>{\setlength\hsize{1.2\hsize}\raggedright}X%
 >{\setlength\hsize{1.2\hsize}\raggedright}X%
 >{\setlength\hsize{0.8\hsize}\raggedright}X%
 >{\setlength\hsize{0.8\hsize}\raggedright}X}
& \multicolumn{1}{>{\bfseries}l}{Step 1}
& \multicolumn{1}{>{\bfseries}l}{Step 2}
& \multicolumn{1}{>{\bfseries}l}{Step 3}  
\tabularnewline \toprule
Eigenspace & 
$U_\mathrm{All}(W_{i^*})$ & 
$U_\mathrm{N}(W_{i^*})$ &
$U_{\mathrm{N}}(W_{i^*},k)$
\tabularnewline \midrule
Data used to train eigenspace &
$\mathcal{D}_\mathrm{All}(W_{i^*})$ &
$\mathcal{D}_\mathrm{N}(W_{i^*})$ &
$\mathcal{D}_{\mathrm{N}}(W_{i^*},k)$
\tabularnewline \midrule
Class 1 data used to find $e_\mathrm{dfes}^1$&
$\mathcal{D}_\mathrm{All}(W_{i^*})$ &
$\mathcal{D}_\mathrm{N}(W_{i^*})$ &
$\mathcal{D}_{\mathrm{N}}(W_{i^*},k)$   
\tabularnewline \midrule
Class 2 data used to find $e_\mathrm{dfes}^2$&
$\mathcal{D}_\mathrm{All}(W_{i})$,$i=1,2,...,L,\mbox{but } i \neq i^*$ &
$\mathcal{D}_\mathrm{NN}(W_{i^*})$ &                    
$\mathcal{D}_{\mathrm{NN}}(W_{i^*},k)$
\tabularnewline \bottomrule
\multicolumn{4}{p{0.75\linewidth}}{Note: $U$ denotes eigenspace, $\mathcal{D}$ denotes data, $\mathrm{All}$ denotes both native and non-native included, $\mathrm{N}$ denotes native, $\mathrm{NN}$ denotes non-native, $L$ denotes the number of different words, $(W_{i^*},k)$ denotes $k^{\mathrm{th}}$ syllable in word $i^*$, and $k=1,2,...,K(i^*)$ where $K(i^*)$ is the number of syllables in word $W(i^*)$}
\end{tabularx}
\end{center}
\end{table}

The main goal of the training process is to find the ``optimal threshold" to detect mispronunciations in the test samples. In each step, since the test samples come from two classes of data which are supposed to be separated, the optimal threshold $T$ should be the one that separates these two classes best. This is done using Bayes rule by finding the optimal threshold to separate two sets of distances $e_\mathrm{dfes}^1$ for class 1 and $e_\mathrm{dfes}^2$ for class 2 data. Let random variables $X_1$ and $X_2$ represent $e_\mathrm{dfes}^1$ and $e_\mathrm{dfes}^2$ and assume they are both Gaussian distributed with prior probabilities $P(\omega_1)$ and $P(\omega_2)$, where $\omega_{1,2}$ denote the classes of these two groups. The classification error can be computed by Equation (\ref{pe}) using Bayes rule

\begin{eqnarray} \label{pe}
	P(\rm{error}) &=& \int_{x^*}^{\infty} p(x \mid \omega_1 ) P(\omega_1 ) \mathrm{d}x + \int_{-\infty}^{x^*} p(x\mid \omega_2 ) P(\omega_2 ) \mathrm{d}x
\end{eqnarray}
where the optimal threshold $x^* \mbox{ or } T$ can be found by computing the discriminant function $g(x)$ at $g(x)=0$ where
\begin{equation} \label{discriminant}
	g(x)=p(x\mid \omega_1 )P(\omega_1 )-p(x\mid \omega_2 )P(\omega_2 )
\end{equation} 

Figure \ref{wdpcamfcc_numerical} illustrates the numerical distribution of $e_\mathrm{dfes}^1$ (top line) and $e_\mathrm{dfes}^2$ (bottom line) as an example of word verification step for the word {\tt tres}. The distances are averaged at 5 repetitions for each speaker. Figure \ref{wdpcamfcc_theoretical} plots the corresponding theoretical Gaussian distribution in order to obtain a more reliable optimal threshold. MFCCs were used as the feature set here. Note that since there are 9 words from class 2 and only 1 word from class 1, distance $e_\mathrm{dfes}^1$ has been repeated 9 times to make them comparable to the $e_\mathrm{dfes}^2$. The optimal threshold $T_d$ can be found by computing the minimum error rate of the theoretical distribution (Gaussian distribution assumed) shown in Figure \ref{wdpcamfcc_theoretical}.  By Bayes rule, the threshold is optimal when it provides the theoretical minimum classification error $P(\rm{error})$ ($0.030\%$ in this case). The $P(\rm{error})$ computed using spectrograms is similar but a little higher ($0.447\%$).

\begin{figure}[h]
\begin{minipage}[t]{0.49\linewidth} 
\centering
\captionsetup{width=0.9\textwidth} 
\includegraphics[height=6.5cm]{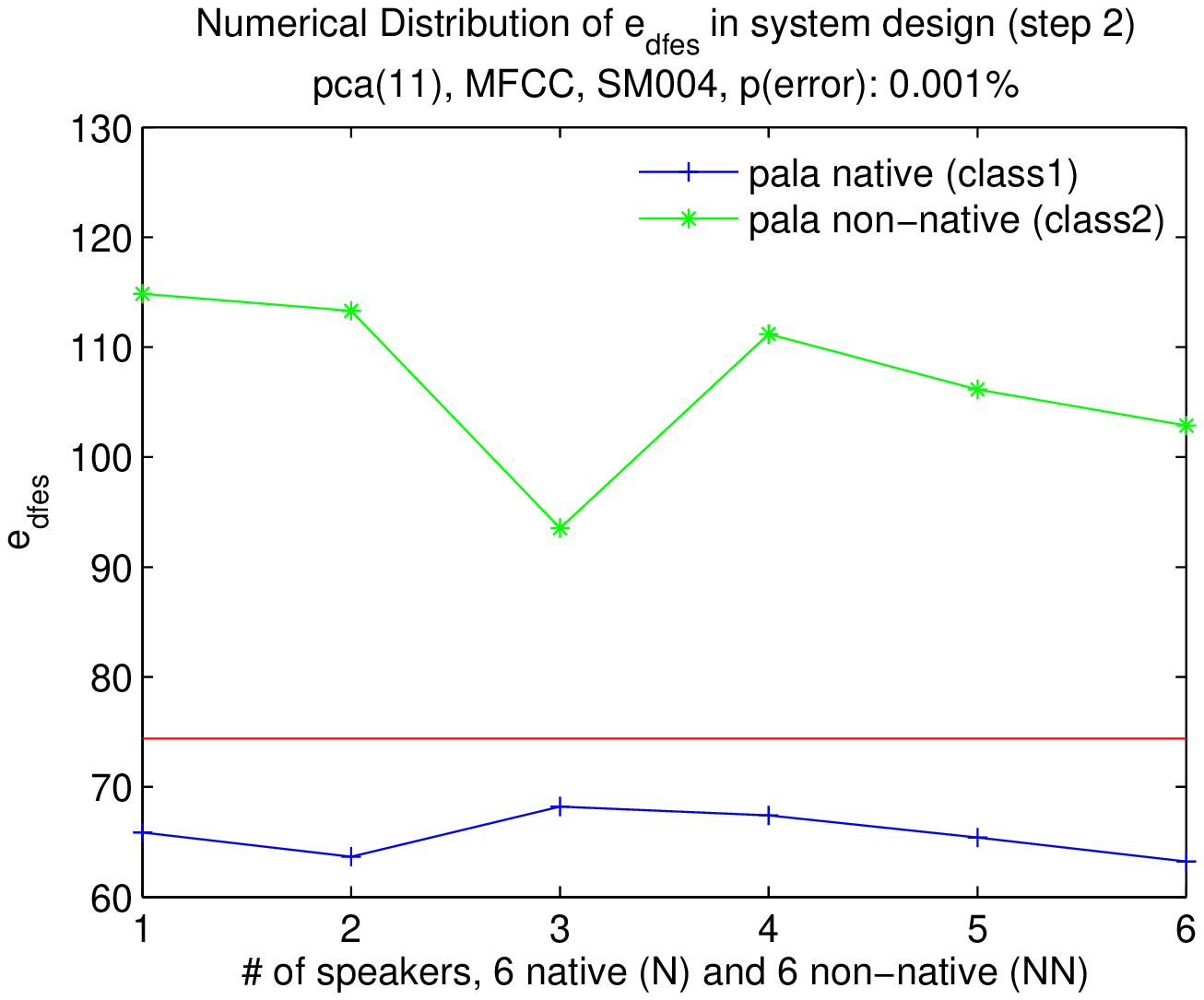}
\caption[width=0.25\textwidth]{Numerical distribution of $e_\mathrm{dfes}$ of the class 1 and class 2 data in N/NN classification (target word: {\tt pala}, feature: MFCCs )}  
\label{wcpca20mfcc_numerical}
\end{minipage}
\hfill
\begin{minipage}[t]{0.49\linewidth}
\centering
\captionsetup{width=0.9\textwidth} 
\includegraphics[height=6.5cm]{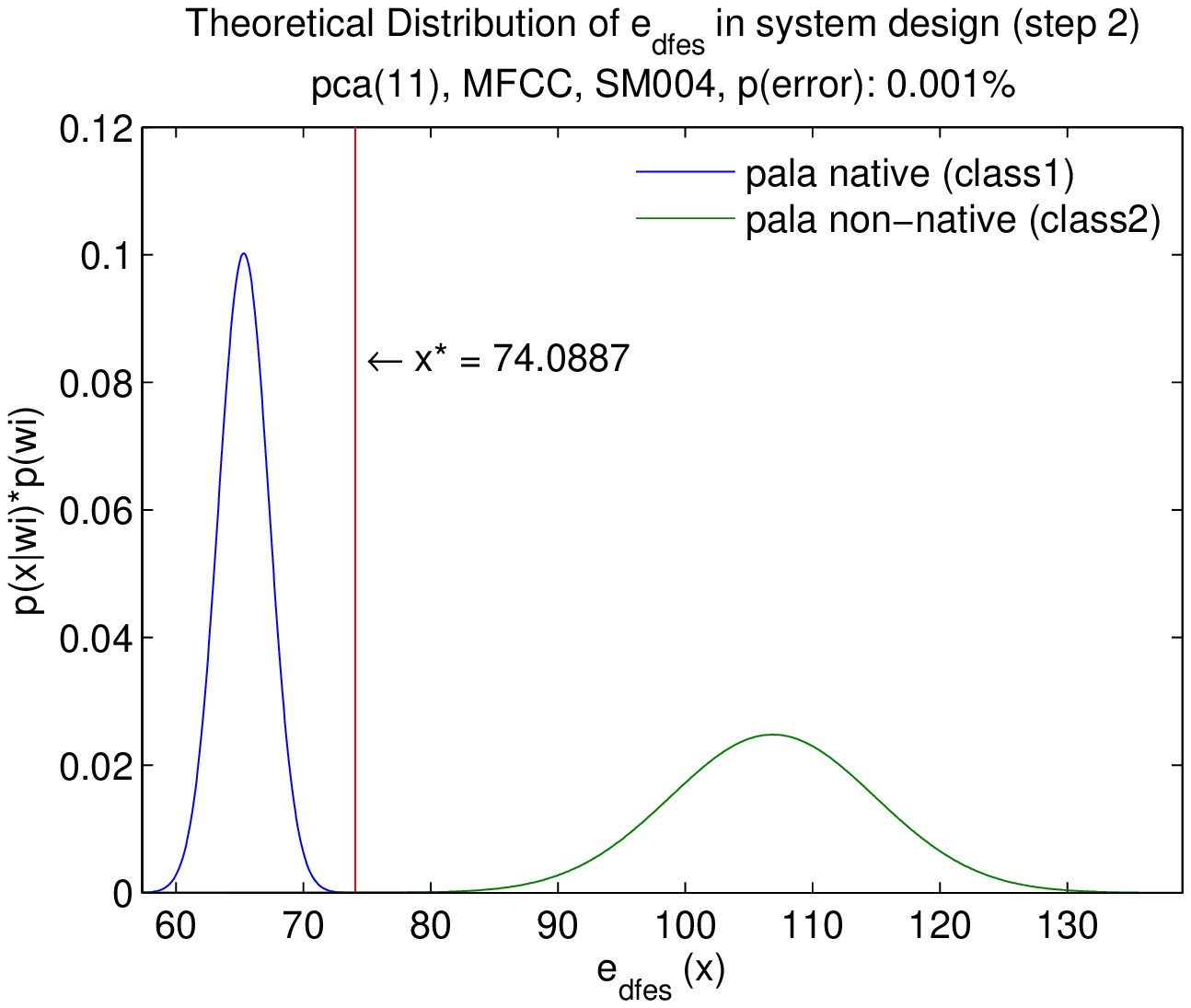}
\caption{Theoretical distribution of $e_\mathrm{dfes}$ of the class 1 and class 2 data in N/NN classification (target word: {\tt pala}, feature: MFCCs )}
\label{wcpca20mfcc_theoretical}
\end{minipage}
\end{figure}

Figure \ref{wcpca20mfcc_numerical} illustrates the native/non-native distribution of data in N/NN classification step for word {\tt pala}. The $P(\rm{error})$ obtained using MFCCs (Figure \ref{wcpca20mfcc_theoretical}) and spectrograms are $0.001\%$ and $11.65\%$ respectively. Experimental results shows that MFCCs are much better than spectrograms in separating data.  

\section{SYSTEM TESTING AND RESULTS}
\label{system testing}

After eigenspace training and detection threshold optimization, the mispronunciation detection system is built up. The following section presents the system testing results in each step.

\subsection{Leave-One-Out Training and Testing}
Because of the relatively small size of the database, the Leave-One-Out (LOO)
method is used for training and testing. Traditionally in LOO, all but one sample is used in training and the left out sample is used for testing. In our case, samples belonging to one speaker (i.e. all 5 repetitions) are left out for testing and all samples belonging to all other speakers are used in system training, which includes the 3 phases discussed in the Section \ref{system training}: eigenspace construction ($U$), 2-class distance measurement ($e_\mathrm{dfes}^1$,$e_\mathrm{dfes}^2$), and detection threshold optimization ($T_d$, $T_c$ and $T_k$).

For example, in the word verification step, the goal is to verify whether or not the ``test word" is the target word $W_{i^*}$. The number of samples available for training and testing in this step is 650 (10 different words $\times$ 13 speakers $\times$ 5 repetitions). Using the LOO method, 5 repetitions of word $W_{i^*}$ from one speaker are left out for testing and the remaining 645 samples are used to train the system and obtain the optimal threshold $T_d$. This is repeated for each speaker. At the end, there are 130 trained system/threshold combinations and 5 test samples for each system to be validated.

In the native/non-native classification, the number of samples available for training and testing is 65 (13 speakers $\times$ 5 repetitions of the target word). Five repetitions from one speaker are left out for testing and the remaining 60 samples are used to train the system and obtain the optimal threshold $T_c$. At the end, there are 13 system/threshold combinations and 5 test samples for each system to be validated. 

\subsection{Results of Word Verification and N/NN Classification}
Compared with the theoretical error rate in Equation (\ref{pe}), the performance of the mispronunciation detection system is measured by the numerical error rate $P_e$
\begin{equation} \label{Pe}
	P_e = \frac{N_{e1}+N_{e2}}{N_1+N_2}~\mbox{,}
\end{equation}
where $N_1$, $N_2$ are the number of test samples from class 1 and 2, and $N_{e1}$, $N_{e2}$ are the number of misclassified samples from each class.

\begin{figure}[h]
\begin{minipage}[t]{0.49\linewidth} 
\centering
\captionsetup{width=0.9\textwidth} 
\includegraphics[height=6.5cm]{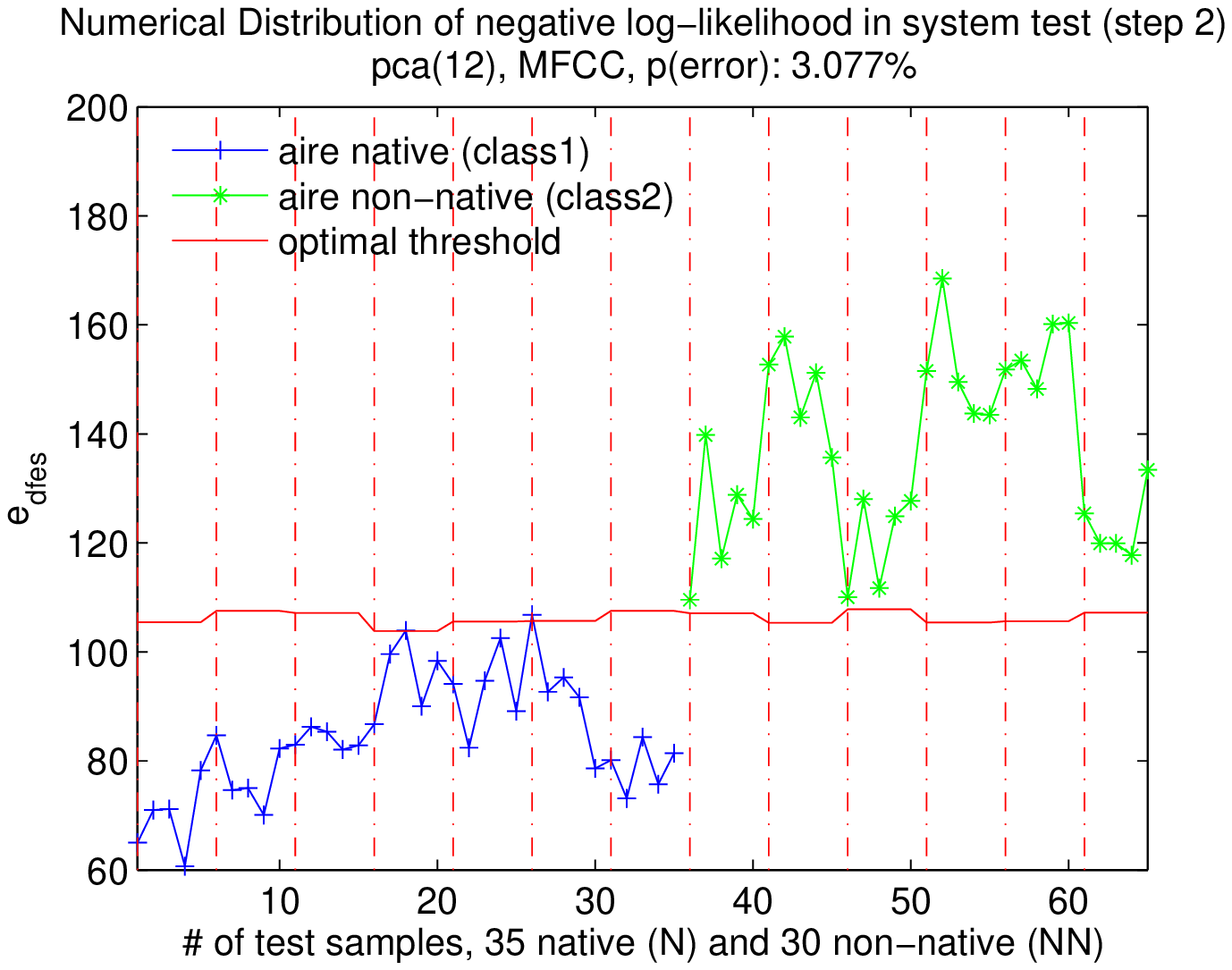}
\caption[width=0.25\textwidth]{Numerical distribution of $e_\mathrm{dfes}$ of the test samples in N/NN classification using PCA (target word: {\tt aire}, feature: MFCCs )}  
\label{wcpca_MFCC_w10-test}
\end{minipage}
\hfill
\begin{minipage}[t]{0.49\linewidth}
\centering
\captionsetup{width=0.9\textwidth} 
\includegraphics[height=6.5cm]{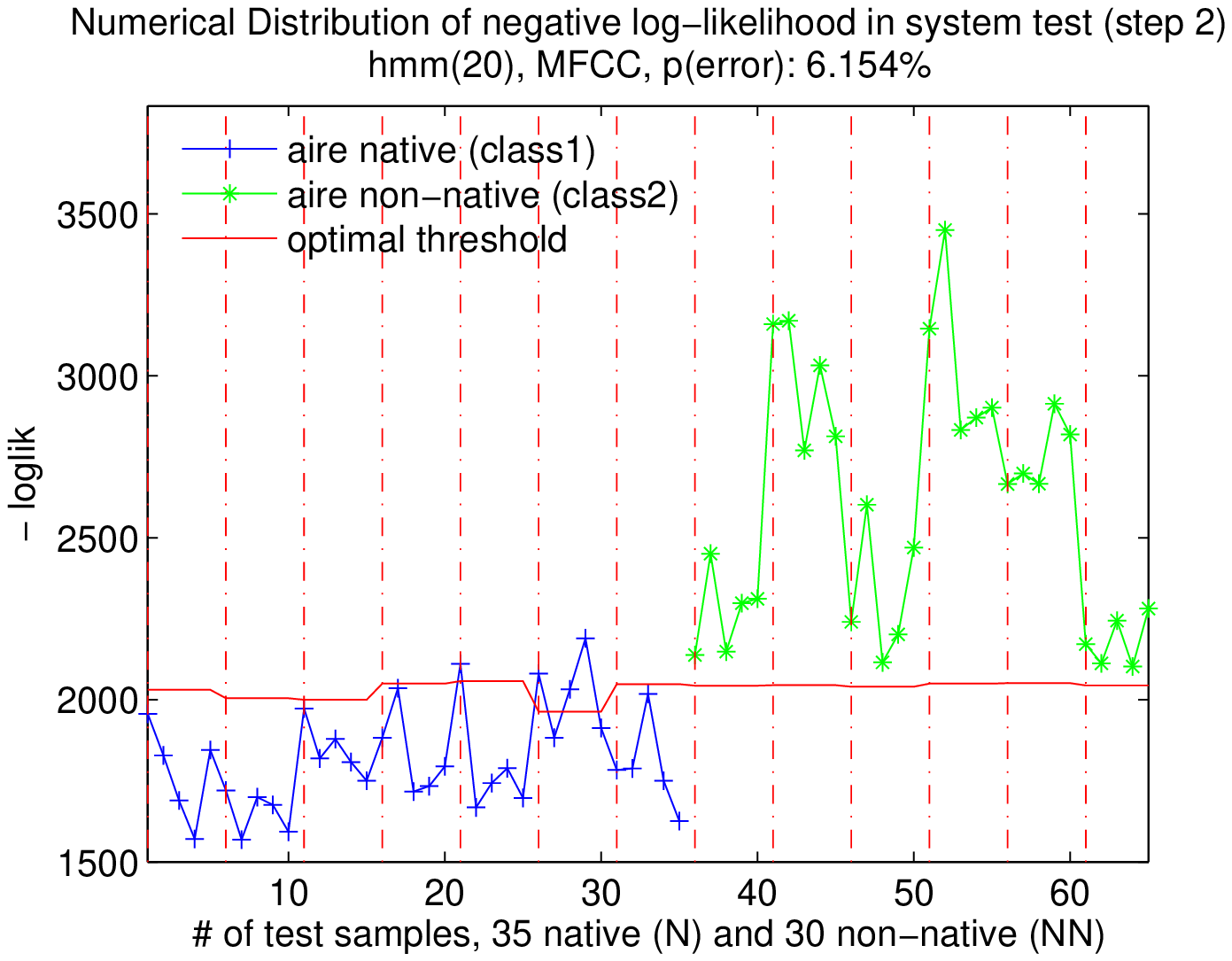}
\caption{Numerical distribution of negative log-likelihood of the test samples in N/NN classification using HMM (target word: {\tt aire}, feature: MFCCs )}
\label{wchmm_MFCC_w10-test}
\end{minipage}
\end{figure}

In word verification, the error rate $P_e$ is always below 3\% and 7\% for MFCCs and spectrograms respectively. In N/NN classification, the performance based on HMMs using MFCCs is also compared along with the PCA method. Figure \ref{wcpca_MFCC_w10-test} and \ref{wchmm_MFCC_w10-test} show the results of the Leave-One-Out method applied to the word {\tt aire} using PCA and HMMs. The 13 columns in the figure represent the 13 speakers of the word {\tt aire}. For each column (speaker), there are 5 samples, which are compared against the threshold. The bottom line corresponds to the 7 native speakers and the top line corresponds to the 6 non-native speakers. The thresholds during each LOO iteration vary slightly because of small differences in the training database during each trial. These variations diminish as the database size increases. For the PCA method using MFCCs, there is 1 sample from both native speaker 4 and 6 in Figure \ref{wcpca_MFCC_w10-test} that are slightly above the threshold and misclassified as non-native and all the rest from both classes are correctly classified. A similar situation is illustrated in Figure \ref{wchmm_MFCC_w10-test} with slightly higher $P_e$. Complete word verification and N/NN classification results of all 10 words are provided in Table \ref{result1}.

In Table \ref{result1}, MFCCs are shown to perform much better than spectrograms using PCA, especially in step 2, N/NN classification. The PCA method performs slightly better on average than HMMs in this data-limited case. However, general parameters of HMMs are used and they are not optimized for this specific system. With respect to the computational cost, training an eigenspace is almost $10^3$ times faster than training HMMs (5 ms vs. 5 s), while in sample testing, the PCA method is also 60-80 times faster (about 4 ms vs. 250 ms) than HMMs depending on the length of the word. 

%

\begin{table}
\begin{small}
\begin{center}
\caption{Error rate $P_e$ in word verification and N/NN classification}
\label{result1}
\renewcommand{\tabcolsep}{1.2pt}
\begin{tabular}{@{} *{15}{c} @{}} \toprule
Steps & Methods & \multicolumn{10}{c}{Words} & Max & Min & Avg.\\
& -Features & $W_1$ & $W_2$ & $W_3$ & $W_4$ & $W_5$ & $W_6$ & $W_7$ & $W_8$ & $W_9$ & $W_{10}$ \\
\midrule
\multirow{2}*{1} & PCA-MFCCs
& 2.04\%   &   0.03\%   &   0.63\%   &   0.07\%   &   0.19\%
& 0.18\%   &   0.37\%   &   1.04\%   &   0.08\%   &   1.43\%
& 2.04\%   &   0.03\%   &   \textbf{0.61\%} \\
\cmidrule(l){2-15}
& PCA-Spec.
& 6.70\%   &   0.45\%   &   1.56\%   &   2.67\%   &   5.40\%
& 3.47\%   &   3.29\%   &   5.24\%   &   2.22\%   &   6.67\%  
& 6.70\%   &   0.45\%   &   \textbf{3.77\%} \\
\midrule
\multirow{3}*{2} & PCA-MFCCs
& 12.31\%  & 10.77\%   & 1.54\%  & 12.31\%  & 7.69\%  & 7.69\% & 6.15\% & 9.23\%  & 4.62\%  & 3.08\%  
& 12.31\%  & 1.54\%    & \textbf{7.54\%} \\ 
\cmidrule(l){2-15}
& \small{PCA-Spec.}  
& 24.62\%  & 16.92\%  & 27.69\%  & 21.54\% & 18.46\%  & 18.46\% & 10.77\% & 18.85\%  & 13.85\%   
& 15.38\%  & 27.69\%  & 10.77\%  & \textbf{18.15\%} \\
\cmidrule(l){2-15}                  
& HMM-MFCCs
& 3.08\%   & 10.77\%  & 6.51\%   & 9.23\%  & 10.77\%  & 12.31\% & 9.23\%  & 7.69\%  & 4.62\%    
& 6.154\%  & 12.31\%  & 3.08\%   & \textbf{8.04\%}  \\
\bottomrule   
\end{tabular}
\end{center}
\end{small}
\end{table}


\subsection{Results of Syllable-Level Mispronunciation Detection}

For syllable-level mispronunciation detection, an approach similar to the N/NN classification is followed, except that it is done at a syllable level. Though in real applications, only samples that are classified as non-native in the N/NN classification would be processed by the 3$^\mathrm{rd}$ step, to make testing results comparable and unbiased, all test samples including native samples are also used here for evaluation.

There is a major difference between this step and the previous two steps. This is because the assumption that two classes of data in eigenspace training and $e_\mathrm{dfes}$ computing are separated is no longer valid, namely, some syllables may be pronounced well enough that can not be used to distinguish native and non-native. If this is the case, the threshold obtained using Bayes rules is biased and moves towards the native class, which dramatically increases the classification error rate $P_e$. Thus, $P_e$ cannot be used to measure the performance of mispronunciation detection. Instead, it serves as the similarity measurement of the syllables pronounced by both native and non-native classes. By dividing the total error rate $P_e$ into False Negative Rate (FNR) and False Positive Rate (FPR) in Equation (\ref{FNR&FPR}), it is easy to find that with a relatively low FNR, higher FPR indicates better pronunciation of the syllable, while lower FPR shows the problematic syllable that one should pay attention to. 

\begin{equation} \label{FNR&FPR}
	\mathrm{FNR} = \frac{N_{e1}}{N_1} ~~ \textrm{ and } ~~
	\mathrm{FPR} = \frac{N_{e2}}{N_2}
\end{equation}

\begin{figure}[h]
\begin{minipage}[t]{0.49\linewidth} 
\centering
\captionsetup{width=0.9\textwidth} 
\includegraphics[height=6.5cm]{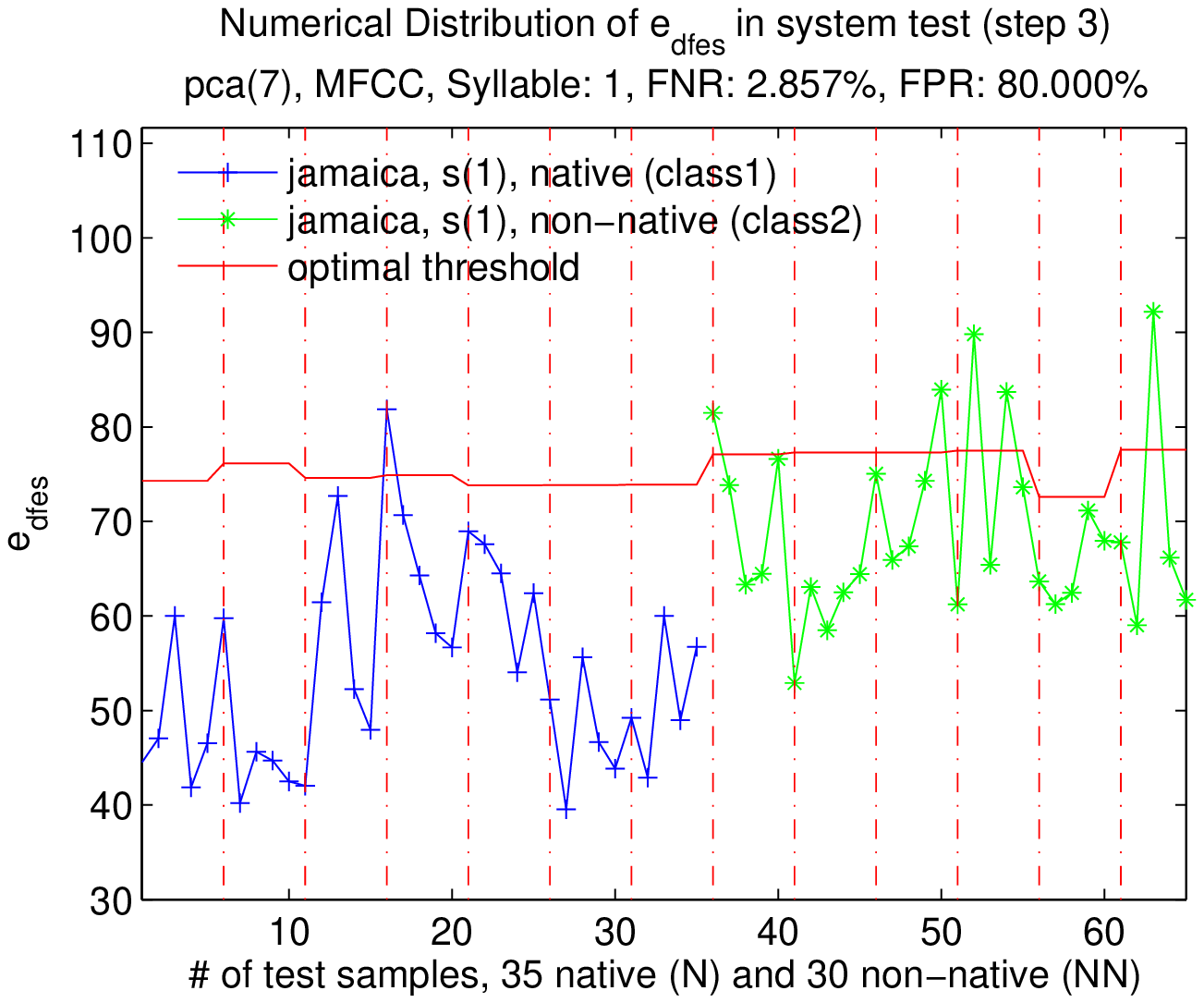}
\caption[width=0.25\textwidth]{Numerical distribution of $e_\mathrm{dfes}$ of the test samples in mispronunciation classification using PCA (target syllable: {\tt jamaica}, /ja/, feature: MFCCs )}  
\label{wcpca_MFCC_w1_p1-test}
\end{minipage}
\hfill
\begin{minipage}[t]{0.49\linewidth}
\centering
\captionsetup{width=0.9\textwidth} 
\includegraphics[height=6.5cm]{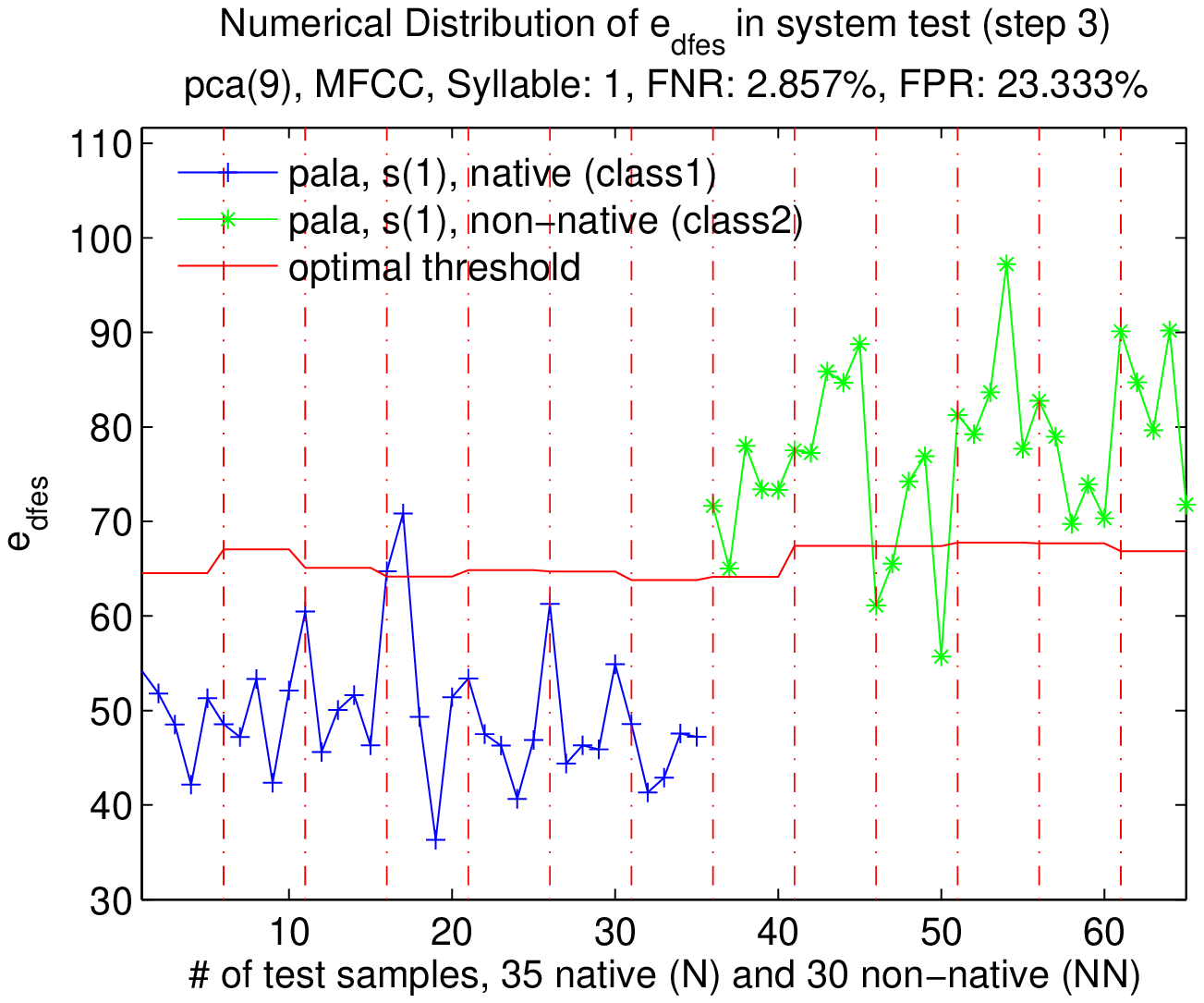}
\caption{Numerical distribution of $e_\mathrm{dfes}$ of the test samples in mispronunciation classification using PCA (target syllable: {\tt pala}, /pa/, feature: MFCCs )}
\label{wcpca_MFCC_w5_p1-test}
\end{minipage}
\end{figure}

Figure \ref{wcpca_MFCC_w1_p1-test} and \ref{wcpca_MFCC_w5_p1-test} illustrate two examples where ``good" (FPR = 80\%) and ``bad" (FPR = 23.3\%) mispronunciations are detected. Table \ref{result2} shows FNR/FPR for each syllable and highlights all $\mathrm{FPR} \leq 30\%$. From Table \ref{result2}, common mispronunciation problems\cite{Hammond} like vowel reduction, aspiration, linkage and stress are successfully detected in the test database. Some of them are listed below:

\begin{itemize}
\item Vowel Reduction: /pa/ in {\tt pala}; /den/ in {\tt accidente}; /ie/ in {\tt arie}
\item Aspiration: /pa/ in {\tt pala}; /puer/ in {\tt puertorriquena}
\item Linkage: /tr/ in {\tt tres}; /cons/ in {\tt construccion}; /na/ in {\tt puertorriquena}
\item Stress: /ci/ in {\tt accidente}; /cons/ and /cion/ in {\tt construccion}
\end{itemize}

\begin{table}
\begin{center}
\caption{FNR and FPR in mispronunciation detection}
\label{result2}
\begin{tabular}{@{} *{2}{lccc} @{}} \toprule
\multicolumn{1}{>{\bfseries}l}{Words} &
\multicolumn{1}{>{\bfseries}l}{Syllables} &
\multicolumn{1}{>{\bfseries}l}{FNR$^*$} &
\multicolumn{1}{>{\bfseries}l}{FPR$^{**}$} &
\multicolumn{1}{>{\bfseries}l}{Words} &
\multicolumn{1}{>{\bfseries}l}{Syllables} &
\multicolumn{1}{>{\bfseries}l}{FNR} &
\multicolumn{1}{>{\bfseries}l}{FPR} 
\tabularnewline \toprule
{\tt jamaica}    &   /ja/    &   2.86\%   &   80.00\%   &   {\tt pala}          &   /pa/    &   2.86\%   & \textbf{23.33\%}  \\
                 &   /mai/   &   0.00\%   &   80.00\%   &                       &   /la/   &   0.00\%   & 80.00\%  \\
\cmidrule(l){6-8}
                 &   /ca/    &  11.43\%   &   53.33\%   &  {\tt accidente}      &   /ac/    &   5.71\%   & 53.33\%  \\
\cmidrule(l){2-4}
{\tt tres}       &   /tr/    &   5.71\%   &   \textbf{16.67\%}   &                       &   /ci/    &   2.86\%   & \textbf{26.67\%}  \\
                 &   /e/     &   0.00\%   &   73.33\%   &                       &   /den/   &   5.71\%   & \textbf{10.00\%}  \\
                 &   /s/     &  11.43\%   &   36.67\%   &                       &   /te/    &   8.57\%   & 66.67\%  \\
\cmidrule(l){2-4} \cmidrule(l){6-8}
{\tt gemelas}    &   /ge/    &   5.71\%   &   40.00\%   & {\tt construccion}    &   /cons/  &   0.00\%   & \textbf{30.00\%}  \\
                 &   /me/    &   2.86\%   &   43.33\%   &                       &   /truc/  &   2.86\%   & 40.00\%  \\
                 &   /la/    &   2.86\%   &   80.00\%   &                       &   /cion/  &   8.57\%   & \textbf{30.00\%}  \\
\cmidrule(l){6-8}
                 &   /s/     &   5.71\%   &   60.00\%   & {\tt puertorriquena}  &   /puer/  &   5.71\%   & \textbf{30.00\%}  \\
\cmidrule(l){2-4}
{\tt hierro}     &   /hie/   &   2.86\%   &   56.67\%   &                       &   /tor/   &   2.86\%   & 80.00\%  \\
                 &   /rro/   &   8.57\%   &   40.00\%   &                       &   /ri/    &   0.00\%   & 70.00\%  \\
\cmidrule(l){2-4}
{\tt torturados} &   /tor/   &   8.57\%   &   \textbf{30.00}\%   &                       &   /que/   &   2.86\%   & 70.00\%  \\
                 &   /tu/    &   2.86\%   &   43.33\%   &                       &   /na/    &   2.86\%   & \textbf{26.67\%}  \\
\cmidrule(l){6-8}
                 &   /ra/    &   8.57\%   &   70.00\%   &   {\tt aire}          &   /ai/    &   2.86\%   & 33.33\%  \\
                 &   /do/    &   17.14\%  &   73.33\%   &                       &   /re/    &   0.00\%   & \textbf{30.00\%}  \\
                 &   /s/     &   8.57\%   &   86.67\%  
\\\bottomrule
\multicolumn{8}{l}{$^*$FNR: False Negative Rate; $^{**}$FPR: False Positive Rate}
\end{tabular}
\end{center}
\end{table}

\section{CONCLUSION AND FUTURE WORK}

The testing results in Section \ref{system testing} show that PCA can be a computationally efficient approach to detect mispronunciations, even when the training and testing database is limited. MFCCs were shown to outperform spectrograms. Compared with HMMs, the PCA method is much faster and achieves comparable results for the database used in this paper. 

For future work, the threshold in each step should be optimized on a larger database to improve robustness. This is especially important for optimizing the threshold in the syllable-level detection, since subtle differences that reside in syllables require more data to differentiate. 

Because of the constraints of the PCA method in data separation (discriminant information may stay in the less significant components), further investigation on the distributions of multiple types of mispronunciations is warranted. Furthermore, hybrid methods based on PCA, LDA (Linear Discriminant Analysis) and ICA (Independent Component Analysis) should also be considered. 

\bibliography{mymanuscript_v07}
\bibliographystyle{spiebib} 

\end{document}